\documentclass[prb,aps]{revtex4}
\usepackage{graphicx}

\newcommand{\beq}{\begin{eqnarray}}
\newcommand{\eeq}{\end{eqnarray}}
\renewcommand{\vec}[1]{{\mathbf{#1}}}
\begin{document}
\draft
%\twocolumn
%\preprint{dvi file made on \today}
%\input epsf.sty
%\input psfig.sty

\title
{Vortex Glass is a Metal: 
Unified Theory of the Magnetic Field and Disorder-Tuned Bose Metals}

\author{ Jiansheng Wu and Philip Phillips}
\affiliation{Loomis Laboratory of Physics,
University of Illinois at Urbana-Champaign,
1100 W.Green St., Urbana, IL., 61801-3080}
\vspace{.05in}

\begin{abstract}
We consider the disordered quantum rotor model in the presence of a magnetic field.  We analyze the transport properties in the vicinity of the multicritical point between the superconductor, phase glass and paramagnetic phases.  We find that the magnetic field leaves metallic transport of bosons in the glassy phase in tact.  In the vicinity of the vicinity of the superconductivity-to-Bose metal transition, the resistitivy turns on as $(H-H_c)^{2}$ with $H_c$.  This functional form is in excellent agreement with the experimentally observed
turn-on of the resistivity in the metallic state in MoGe, namely $R\approx R_c(H-H_c)^\mu$, $1<\mu<3$.  The metallic state is also shown to presist in three spatial dimensions. In addition, we also show that the metallic state remains intact in the presence of Ohmic dissipation in spite of recent claims to the contrary.
As the phase glass in $d=3$ is identical to the vortex glass,
we conclude that the vortex glass is, in actuality, a metal rather than a superconductor at $T=0$.  Our analysis unifies the recent experiments on vortex glass systems in which the linear resistivity remained non-zero below the putative vortex glass transition and the experiments on thin films in which a metallic phase has been observed to disrupt the direct transition from a superconductor to an insulator.
\end{abstract}

\maketitle

\section{Introduction}

When superconductivity is quenched in thin metal alloy films, a metallic\cite{jaeger,mooij,yaz,mason} rather than an insulating phase ensues.   The subsequent metallic phase is robust as it
 exists over a wide range of magnetic field and film thickness (disorder).
Regardless of the statistics of the charge carriers,
a $T=0$ metallic phase in two dimensions does not mesh easily
with conventional wisdom.  In the case of bosons, the uncertainty principle dictates that in the absence of some new state, bosons either condense or
localize as in conventional treatments\cite{mpaf} of the insulator-superconductor problem.  Further, non-interacting electrons\cite{abrahams} carry no curent in two dimensions in the presence of disorder.  Precisely what the new exotic metallic state is is hotly debated\cite{denis,fishernew}.
While many proposals have been made\cite{denis}, only the phase glass model\cite{denis,pglass,qstiff} has been demonstrated through direct calculation to have a finite resistivity at $T=0$ {\bf in the presence of disorder}.  In this model, the charge carriers are strictly bosonic.  Metallic transport arises from the coupling of the bosonic degrees of freedom to the low-lying excitations of the glassy phase.  The density\cite{pglass,rsy,huse,bm} of such low-lying excitations scales as $|\omega|$ and hence
gives rise to a dynamical exponent of $z=2$. Indeed, other models have been proposed which could give rise to $z=2$ dynamics.  For example, a popular view is that the metallic state for the bosons is driven by dissipation arising from
normal electronic excitations\cite{chakdiss,chakravarty}.  Such fermionic excitations are typically modeled by a dissipative Ohmic heat bath. However, bosons described by an Ohmic propagator of the form,
\beq
G_0=(k^2+\eta|\omega_n|+m^2)^{-1},
\eeq
still only exhibit either superconducting or insulating phases\cite{wagen,dp}. Aside from providing a dissipative environment, normal electronic excitations could carry the current in the metallic state.  As the normal state of the materials of interest, MoGe for example, is well described by Fermi liquid theory, the localization principle, which precludes a metallic state for non-interacting electrons in two dimensions\cite{abrahams}, should be operative.  Hence, while normal electronic excitations are undoubtedly present above the superconducting state, invoking\cite{kapnew} them to explain the origin of the metallic state is a non-starter.
Additional fermionic models have been suggested\cite{fishernew} in which bosonic vortices are attached to a strongly fluctuating gauge field which transmutates the statistics.  On phenomenological grounds, the resultant fermionic vortices were assumed to diffuse and hence constitute a gapless liquid.   While flux attachment schemes have been highly successful in describing gapped phases\cite{jain}, such threadings are problematic
in gapless states\cite{hlr} as the fluctuations of the gauge field are relevant\cite{larkin}. 

In the phase glass model, fermionic excitations need not be invoked because the dissipation is self-generated from the glassy environment.  The resultant bosonic propagator (derived in the Appendix A) is of the form,
\beq\label{prop}
G^{(0)}_{ab}(\vec k,\omega_n)=G_0 (\vec k,\omega_n)\delta_{ab}+
\beta G_0^2(\vec k,\omega_n)q\delta_{\omega_n,0}
\eeq
where $a$ and $b$ are the replica indices, $q$ is the Edwards-Anderson order parameter for the glassy phase and $\omega_n$ are the Matsubara frequencies. The first term is the standard bosonic propagator.  However, the second term arises entirely from the glassy landscape.  Although Eq. (\ref{prop}) can be derived explicitly, it also follows directly from dimensional reduction. That the second term contains the square of the bare bosonic propagator, Zinn-Justin\cite{zinjust} surmised holds profound consequences for
disordered systems.  Indeed this is true.  It is from this term that the metallic state arises.

To constitute a complete theory of the metallic state that disrupts the insulator-superconductor transition, the phase glass must be applicable to systems in the presence of a magnetic field as well.  It is precisely this question that we address here. We show that the metallic state persists in the presence of a
magnetic field.  In the weak-field limit, we find that the dominant role of the field is to renormalize
the inverse correlation length. The low-field limit of the resistivity is described by a power law of the form $R_c(H-H_c)^{2z\nu}$, where $z=2$ and 
$\nu=1/2$ at the mean-field level. This functional form is in excellent agreement with the experimentally observed turn-on of the resistivity in MoGe\cite{masonthesis}.  We also find that the metallic state survives in 3D.  In 3D the relevant comparison is with the vortex glass\cite{ffh} which has been
argued extensively to be a superconducting phase.  We find instead that this state is metallic\cite{zimanyi,ybco,scaling}.

\subsection{Phase Glass Overview}

We begin by summarizing the key elements of the zero-field theory we will need when we consider the effects of a non-zero magnetic field.  The phase glass arises naturally in the context of the disordered quantum rotor model
\beq\label{qrm}
H=-E_C\sum_i\left(\frac{\partial}{\partial\theta_i}\right)^2-
\sum_{\langle i,j\rangle} J_{ij}\cos(\theta_i-\theta_j),
\eeq
with random Josephson couplings $J_{ij}$ but fixed
on-site energies, $E_C$.  The phase of each superconducting island
is $\theta_i$.  Note that additional on-site disorder of the form
$iv_j\partial /\partial \theta_j$ results in the equivalent particle-hole symmetric
field theory provided that
the distribution of on-site energies has zero mean.  For the on-site energies,
the non-zero
mean case is irrelevant here as this corresponds to a density-driven
insulator-superconductor transition (IST)\cite{sachbook}.

If the Josephson couplings are chosen from a distribution
with zero mean, only two phases are possible: 1) a glass arising from the distribtion of positive and negative $J_{ij}'s$ and 2) a disordered paramagnetic state.  A superconducting phase obtains if the distribution
\beq\label{distj}
P(J_{ij})=\frac{1}{\sqrt{2\pi
J^2}}\exp{\left[-\frac{(J_{ij}-J_0)^2}{2J^2}\right]}
\eeq
of $J_{ij}'s$ has non-zero mean, $J_0$, and $J$ the variance.
 To distinguish between the phases, it is expedient
to introduce the set of variables ${\bf S}_i=(\cos\theta_i,
\sin\theta_i)$
which allows us to recast the interaction term
in the random Josephson Hamiltonian as a spin problem
with random magnetic interactions,
 $\sum_{\langle i,j\rangle}J_{ij}{\bf S}_i \cdot {\bf S}_j$.
Let $\langle ...\rangle$ and $[...]$ represent
averages over the thermal degrees of freedom
and over the disorder, respectively.  Integrating over the random interactions
will introduce two auxillary fields
\beq
 Q_{\mu\nu}^{ab}(\vec k,\vec k',\tau,\tau')=\langle
S_\mu^a(\vec k,\tau)S_\nu^b(\vec k',\tau')\rangle
\eeq
and $\Psi^a_\mu(\vec k,\tau)=\langle S^a_\mu(\vec k,\tau)\rangle$,
respectively. The
superscripts represent the replica indices. A non-zero value of
$\Psi^a_\mu(\vec k,\tau)$ implies phase ordering of
the charge $2e$ degrees of freedom.
For quantum spin
glasses, it is the diagonal elements of the Q-matrix
$D(\tau-\tau')=\lim_{n\rightarrow 0}\frac{1}{Mn}\langle
Q^{aa}_{\mu\mu}(\vec k,\vec k',\tau,\tau')\rangle$ in the limit that
$|\tau-\tau'|\rightarrow\infty$ that serve as the effective
Edwards-Anderson spin-glass order parameter\cite{rsy,huse,bm}
within Landau theory. The essential aspect of the quantum rotor spin glass is that the saddle point solution
for the corresponding action is minimied by a solution of the form
\beq\label{qm}
Q_{\mu\nu}^{ab}(\vec k,\omega_1,\omega_2)&=&\beta(2\pi)^d\delta^d(k)
\delta_{\mu\nu}\left[
D(\omega_1)
\delta_{\omega_1+\omega_2,0}\delta_{ab}\right.\nonumber\\
&&\left.+\beta\delta_{\omega_1,0}\delta_{\omega_2,0}q^{ab}
\right]
 \eeq
where the diagonal elements are given by
\beq
D(\omega)=-\sqrt{\omega^2+\Delta^2}/\kappa,
\eeq
with $\kappa$ a coupling constant in the Landau free energy for the spin glass.
The diagonal elements of the $Q$-matrices describe the excitation spectrum.
Throughout the glassy phase, $\Delta=0$ and hence the spectrum is ungapped and given by
$D(\omega)=-|\omega|/\kappa$. The linear dependence on $|\omega|$ arises
because the correlation function $Q^{aa}_{\mu\mu}(\tau)$ decays\cite{rsy,huse,bm} as
 $\tau^{-2}$.
 This dependence results in a fundamental change in the dynamical critical exponent from
$z=1$ to $z=2$. In a recent treatment of the disordered quantum rotor problem,
Ikeda\cite{ikeda} included an Ohmic dissipative $|\omega|$ term at the level of the Hamiltonian.  He has found that the excitation spectrum
in the spin-glass order parameter scales as $|\omega|^{1/2}$.  With the $|\omega|^{1/2}$ ansatz, Ikeda\cite{ikeda} has found that the resultant phase is a superconductor.
We shown in Appendix B, however, that this claim is without merit. Namely, even if $|\omega|$ dissipation is put in by hand at the outset, the universality class of the problem does not change.  Such a contribution simply renormalizes the coefficient of $|\omega|$.

To analyze the bosonic conductivity, we
focus\cite{pglass} on the part of the free energy
\beq\label{fen}
&&{\cal F}[\Psi]=
\sum_{a,\mu, k,\omega_n}(k^2+\omega_n^2+m^2)
|\Psi_\mu^a(\vec k,\omega_n)|^2 \nonumber\\
&&-\frac{1}{\kappa t}\int d^d x\int d\tau_1
d\tau_2\sum_{a,b,\mu, \nu}\Psi_\mu^a(x,\tau_1)
\Psi^b_\nu(x,\tau_2)
Q_{\mu\nu}^{ab}(x,\tau_1,\tau_2)\nonumber\\
&&+U\int d\tau\sum_{a,\mu}\left[\Psi_\mu^a(x,\tau)\Psi_\mu^a(x,\tau)\right]^2
\eeq
governed by the fluctuations of the superconducting order parameter.  The coupling between the spin-glass and
bosonic degrees of freedom fundamentally changes the dynamics as can be seen by substituting the saddle-point
solution for the Q-matrices into Eq. (\ref{fen}).  The resulting expression for the free energy
\beq\label{fgauss}
{\cal F_{\rm gauss}}&=&\sum_{a,\vec k,\omega_n}
(k^2+\omega_n^2+\eta |\omega_n|+m^2)
|\psi^a(\vec k,\omega_n)|^2 \nonumber\\
&&-\beta q\sum_{a,b,\vec k,\omega_n} \delta_{\omega_n,0}
\psi^a(\vec k,\omega_n)[\psi^b(\vec k,\omega_n)]^{\ast},
\eeq
reflects the damped dynamics introduced by the glass degrees of freedom. As remarked earlier,
the $|\omega|$ term dominates giving rise to $z=2$ dynamics.  Certainly, dissipative models
with $|\omega|$ dynamics\cite{chakdiss} put in by hand have been studied explicitly in the context of the insulator-superconductor
transition partly in the hope of finding new phases\cite{chakravarty}.  However, direct calculations\cite{wagen}
of the conductivity reveal that $|\omega|$ dynamics alone cannot give rise to a metallic
phase.   What is new here is that the $|\omega|$ term is not put in by hand but arises naturally from the coupling between the glassy dynamics of the phase and the charge degrees of freedom.   Further,
as is evident from Eq. (\ref{prop}), the Gaussian propagator changes fundamentally as a result of the ulrametric properties of the
glassy fields, $q^{\rm ab}$.  Substitution of Eq. (\ref{prop}) into the Kubo formula for the conductivity\cite{pglass}
\beq\label{cond1}
\sigma(\omega=0,T\rightarrow 0)=\frac{4e^2}{\hbar}\frac{q\eta}{3 m^4}
\eeq
demonstrates that both the $|\omega|$ dynamics and the glassy landscape conspire to yield a non-zero bosonic
conductivity at $T=0$.  This expression crosses over smoothly to the $\sigma=\infty$ conductivity
in the superconducting phase ($m=0$) to the vanishing conductivity of the insulating state ($q=0$).
Hence, the quantum rotor model in which the disordered Josephson couplings possess a non-zero mean describes a metal, insulator and a superconductor,

\subsection{Magnetic Field: Summary of Findings}

In this paper, we extend the formalism summarised above to include a magnetic field. Since disorder is also included, our results in 3D are relevant to the vortex glass state\cite{ffh}.  In the weak field limit, we find that the transport properties are well approximated by replacing $m^2$ in the conductivity by
\beq
\tilde{m}^2=m^2+\frac12 m_H^2
\eeq
where $m_H^2=\frac{2e^\ast H}{h}=2/\ell^2$ is the inverse square of the magnetic length ($\ell$) in units of $c=1$.  Physically, this result is transparent as the predominant role of the magnetic field
is to provide a confining potential for the electrons on a scale set by $1/\sqrt{m_H^2}$. The new condition for the onset of superconductivity
is
\beq
m^2+\frac12 m_H^2=0\quad{\rm superconductivity \quad criterion}
\eeq
which has a non-trivial solution because $m^2<0$ demarcates the superconducting phase.  Hence, we find  quite generally that metallic transport survives in the presence of a magnetic field.  Further, we find that metallic transport persists even in d=3 where the vortex glass phase has been proposed to exist in the presence of disorder\cite{ffh}.  Because of the equivalence of the phase glass to a vortex glass in the presence of magnetic field, our results imply that the linear conductivity should remain finite in the vortex glass.  These results are discussed extensively in the context of the insulator-superconductor transition and the recent experiments that indicate that the vortex glass state is not accompanied\cite{ybco,scaling} by a vanishing of the linear
 I-V characteristics.

\section{Phase Glass Conductivity in a Magnetic Field}

In the presence of a magnetic field, the $O(2)$ quantum rotor model becomes
\beq
H=-E_C\sum_i\left(\frac{\partial}{\partial\theta_i}\right)^2-
\sum_{\langle i,j\rangle} J_{ij}\cos(\theta_i-\theta_j-A_{ij}),
\eeq
where $A_{ij}=(e^* /\hbar) \int_i^j {\vec A}\cdot d{\vec l}$ ($e^* =2e$).
The Josephson couplings are assumed to be random and governed by the
distribution Eq. (\ref{distj}).  In the current work, we treat $A$ to be a static time-independent magnetic field as this is the most experimentally relevant situation\cite{magclaim}. For a
random system, the technique for treating disorder is now standard\cite{dp2}:
1) replicate the partition function,
2) perform the average over disorder and 3) introduce the appropriate
fields to decouple the interacting terms that arise.  As the corresponding
action has been detailed previously\cite{dp2,rsy}, we will provide
additional steps that are necessary to determine how the electromagnetic
gauge couples to the spin glass order parameter. We write the replicated
partition function as
\beq\label{zn}
\overline{Z^n}=\int {\cal D}\theta_i{\cal D}J_{ij}e^{-S}
\eeq
where the Euclidean action is given by
\begin{widetext}
\beq
S=\int_0^\beta d\tau\left\{\sum_{i,a}\frac{1}{4E_C}\left(\frac{\partial
\theta_i^a (\tau)}{\partial\tau}\right)^2  -\sum_a\sum_{\langle ij\rangle}
J_{ij}\cos\left[\theta_i^a (\tau) -\theta_j^a (\tau) -
A_{ij}(\tau)\right]\right\}
\eeq
\end{widetext}
where the superscript $a$ represents the replica index.
The integration over $J_{ij}$
 in Eq. (\ref{zn}) results in the effective action,
\begin{widetext}
\beq
S_{\rm  eff}&=&\int_0^\beta d\tau\sum_{i,a}\frac{1}{4E_C}\left(
\frac{\partial\theta_i^a}{\partial\tau}\right)^2\nonumber\\
&&+\frac{J^2}{2}\sum_{\rm a,b}\sum_{\langle ij\rangle}
\int_0^\beta\int_0^\beta d\tau d\tau'\frac{1}{4}\sum_{\alpha=+1,-1}
\exp\left\{i\left[\theta_i^{a}(\tau)-\alpha\theta_i^b(\tau')-
\left(\theta_j^{a}(\tau)-\alpha\theta_j^b(\tau')\right)-\left(A_{ij}(\tau)-
\alpha A_{ij}(\tau')\right)\right]\right\}\nonumber\\
&& +c.c.\nonumber\\
&&-\frac{J_0}{2}\sum_\alpha\sum_{ij}\int_0^\beta d\tau \exp\left(\theta^a_i-\theta_j^b(\tau)-A_{ij}\right)+c.c.
\eeq
\end{widetext}
with $\alpha=+1,-1$.  As a result of the sum over $\alpha$, we see that the
vector potential enters both symmetrically and anti-symmetrically.
As in the zero-field case, we introduce the two-component spinor, $S^a(\tau)$
and the auxillary field defined in Eq. (\ref{qm}) to describe the spin-glass transition.
The remaining steps involve performing the cumulant expansion
and taking the continuum limit.  The final action can be separated into
the local and gradient parts:
\beq
S_{\rm eff}=S_{\rm loc} +S_{\rm gr}
\eeq
where the local part
\begin{widetext}
\begin{eqnarray} \label{action}
S_{\rm loc}& = & \int d^dx\left\{\frac{1}{\kappa}\int
d\tau\sum_a
\left(r+\frac{\partial}{\partial\tau_1}\frac{\partial}{\partial\tau_2}
\right)Q_{\mu\mu}^{aa}(\vec x ,\tau_1,\tau_2)|_{\tau_1=\tau_2=\tau}
\right.\nonumber\\
&&-\frac{\kappa}{3}\int d\tau_1 d\tau_2
d\tau_3\sum_{a,b,c}Q^{ab}_{\mu\nu}(\vec x ,\tau_1,\tau_2)
Q^{bc}_{\nu\rho}(\vec x ,\tau_2,\tau_3)
Q^{ca}_{\rho\mu}(\vec x ,\tau_3,\tau_1)\nonumber\\
&&\left.+\int d^dx\frac12\int
d\tau\sum_a\left[uQ^{aa}_{\mu\nu}(\vec x ,\tau,\tau)
Q^{aa}_{\mu\nu}(\vec x ,\tau,\tau)
+vQ^{aa}_{\mu\mu}(\vec x ,\tau,\tau)Q^{aa}_{\nu\nu}(\vec x ,\tau,\tau)
\right]\right\}\nonumber\\
&&-\frac{y_1}{6t}\int d^d x\int
d\tau_1d\tau_2\sum_{a,b}\left[Q^{ab}_{\mu\nu}(\vec x ,\tau_1,\tau_2)
\right]^4\nonumber\\
&&+\int d^dx\int
d\tau\sum_a\frac{\partial}{\partial\tau_1}\Psi_\mu^a(x,\tau_1)
\frac{\partial}{\partial\tau_2}\Psi_\mu^a(x,\tau_2)|_{\tau_1=\tau_2=\tau}
+\frac{\zeta}{2}\int
d\tau\sum_a\left[\Psi_\mu^a(x,\tau)\Psi_\mu^a(x,\tau)\right]^2\nonumber\\
&&-\frac{1}{\kappa t g}\int d^d x\int d\tau_1
d\tau_2\sum_{a,b}\Psi_\mu^a(x,\tau_1)
\Psi^b_\nu(x,\tau_2)Q_{\mu\nu}^{ab}(x,\tau_1,\tau_2)+\cdots
\end{eqnarray}
\end{widetext}
contains a spin-glass part that is identical to that of Read, Sachdev and Ye\cite{rsy}
, a superconducting part derived previously
and a gradient part
\beq\label{grad}
S_{\rm gr}&=&\int d^dx\int_0^\beta d\tau_1 d\tau_2\sum_{a,b}
\left|\left(\nabla-\frac{ie^*}{\hbar}
\vec A(\vec x,\tau_1)+\frac{ie^*}{\hbar}\vec A(\vec x ,\tau_2)\right)
Q^{ab}_{-}(\vec x ,\tau_1,\tau_2)\right |^2\nonumber\\
&&+\int d^dx\int_0^\beta d\tau_1 d\tau_2\sum_{a,b}
\left| \left(\nabla-\frac{ie^*}{\hbar}
\vec A(\vec x,\tau_1)-\frac{ie^*}{\hbar}\vec A(\vec x ,\tau_2)\right)
Q^{ab}_{+}(\vec x ,\tau_1,\tau_2) \right|^2\nonumber\\
&&+\frac{1}{2g}\int d^dx\left\{ d\tau \sum_a\Psi_\mu^a(x,\tau)
\left[-\gamma-\left(\nabla-\frac{i e^\ast}{c\hbar}A(x,\tau)\right)\right]\Psi_\mu^a(x,\tau)\right\}
\eeq
in which  the vector potential couples both symmetrically and asymmetrically
to combinations of the $Q-$matrices of the same parity as well as to the bosonic part of the order parameter.
Using the fact that $Q^{ab}_{\pm}(\tau_1,\tau_2) \sim
\langle \exp\left[ i(\theta_i^a(\tau)
\pm\theta_i^b(\tau'))\right]\rangle$, we write
the parity combinations of the $Q-$matrices as follows
\beq\label{qlin}
Q^{ab}_{\pm}(\vec x , \tau_1,\tau_2)=
\frac12\left[ Q_{11}^{ab}(\vec x ,\tau_1,\tau_2)\mp
Q_{22}^{ab}(\vec x ,\tau_1,\tau_2)\right]
+\frac{i}{2}\left[Q_{12}^{ab}(\vec x ,\tau_1,\tau_2)\pm
Q^{ab}_{21}(\vec x ,\tau_1,\tau_2)\right].
\eeq
It is evident that the vector potential enters in a non-time translationally invariant manner.  This is a direct consequence of the fact that the $Q-$matrices
themselves are a function of two independent times, not simply the difference
of $t_1-t_2$.

\subsection{Transport in 2D}

Our immediate focus is on the bosonic fluctuation conductivity in the glass phase. To this end, it suffices to focus on the $\psi$-dependent part of the action:
\beq
S_{\psi}&=&\int d^dx\left\{ d\tau \sum_a\Psi_\mu^a(x,\tau)
\left[-\gamma-\left(\nabla-\frac{i e^\ast}{c\hbar}A(x,\tau)\right)\right]\Psi_\mu^a(x,\tau)\right\}\nonumber\\
&&+\int d^dx\int
d\tau\sum_a\frac{\partial}{\partial\tau_1}\Psi_\mu^a(x,\tau_1)
\frac{\partial}{\partial\tau_2}\Psi_\mu^a(x,\tau_2)|_{\tau_1=\tau_2=\tau}
+\frac{\zeta}{2}\int
d\tau\sum_a\left[\Psi_\mu^a(x,\tau)\Psi_\mu^a(x,\tau)\right]^2\nonumber\\
&&-\frac{1}{\kappa t g}\int d^d x\int d\tau_1
d\tau_2\sum_{a,b}\Psi_\mu^a(x,\tau_1)
\Psi^b_\nu(x,\tau_2)Q_{\mu\nu}^{ab}(x,\tau_1,\tau_2).
\eeq
A precursor to the final result can be obtained by assuming that $\psi$ is independent of both space and time.  In this case, the only vector-potential dependent terms that survive in the total action are those arising from the pure diamagnetic term, $A^2$, in the bosonic part of the action.  For the field along the $z-$axis, we write the vector potential as
\beq
A=\frac12(\vec B\times \vec r).
\eeq
Consequently, the integral of the diamagnetic term over a rectangle with dimensions $L_1\times L_2$ yields $(L_1^2+L_2^2)/3\ell^2$.  This result can be combined into the gradient part of the action for the bosonic degrees of freedom simply by redifining the mass term as
\beq
\gamma\rightarrow\gamma+\frac23(L_1^2+L_2^2)/\ell^2\equiv \tilde{\gamma}.
\eeq
At mean field
\beq
\psi^2=\frac{1}{\xi}(\gamma-\Delta)
\eeq
provides a non-trivial phase boundary for the superconducting state.  In the presence of a magnetic field, our preliminary considerations yield to a modified
boundary
\beq
\tilde{\gamma}\ge\Delta.
\eeq
Because $\tilde{\gamma}>\gamma$, we find that the presence of magnetic field suppresses the superconducting region.  Because $\gamma$ acts as the mass term for the superconducting order,  an intimate connection
\beq
\gamma=\frac{2J_0 d-E_C}{2J_0 d a^2}
\eeq
exists with the basic coupling constants, $E_C$ and $J_0$.  Here we have adopted the mean-field expression\cite{doniach} with $a$ is the distance between the superconducting islands and $d$ is the spatial dimension.  Use of $\tilde{\gamma}$ in this expression results in a mean-field expression for the critical field.  Hence, the renormalization of the mass by the inverse magnetic length
has immediate physical consequences. For example, by simply replacing m by $\tilde{\gamma}$, we obtain the leading magnetic-field
dependence of the conductivity. A more rigorous treatment
will show that this result is almost correct.

A correct treatment of the conductivity requires that we relax the
constraint that $\psi$ be spatially independent.  In this case, we
need to expand $\psi$ ina basis of Landau levels. To this end, we
define \beq\label{psi2d} \psi^a(l,x,y,p_y)=e^{ip_y y}\phi_l(x)
C^a(l,p_y) \eeq where the $\phi_l(x)$ are the Hermite polynomials.
Within the Landau gauge, the gradient part of the action requires
the solution to a Landau level problem of the form, \beq
\left[-\frac{\partial^2}{\partial x^2}-\frac{\partial^2}{\partial
y^2}+ \frac{1}{2\ell^2}p_y\frac{\partial}{\partial
x}+\frac{m_H^4}{4}x^2\right]=\left(l+\frac12\right)m_H^2\phi_l(x).
\eeq
 Expanding the free energy in the Landau level basis results in the replacement of momentum summations with
sums over Landau levels.
The resultant free energy
\beq
F_{\psi}&=&\sum_{a,l,p_y,\omega_m}\left[(l+\frac12)m_H^2+m^2+\omega_m^2+\eta|\omega_m|\right]|C^a(l,p_y,\omega_m)|^2\nonumber\\
&&+\frac{Um_H^2}{4\pi\beta}\int
d\tau\sum_{a,p_y,l,\omega_m}|C(l,p_y,\omega_m)|^4+\beta
q\delta_{\omega_m,0}C^{a\ast}(n,p_y,\omega_m)C^b(l,p_y,\omega_m)
\eeq can be written entirely in terms of the expansion coefficients
$C^a(l,p_y,\omega_m)$ once the Landau level problem is solved. Our
expression for the free energy lays plain that the naive association
of the mass with $m^2+m_H^2/2$ is not entirely correct as an
$l$-dependent term exists as well.  The corresponding Gaussian
propagator is of the form, \beq\label{prop2}
G_{0}^{ab}(l,\omega_n)=G_0 (l,\omega_n)\delta_{ab}+ \beta
G_0^2(l,\omega_n)q\delta_{\omega_n,0} \eeq in the $n\rightarrow 0$
(zero replica) limit with $G_0 (l,\omega_n)=(
m_H^2(l+1/2)+\omega_n^2+\eta|\omega|+\tilde{m}^2)^{-1}$. To
calculate the conductivity, we add to the vector potential a small
perturbation $\delta A(r,t)$ which we use as a source to calculate
the current and the subsequent conductivity which according to the
Kubo formula \beq \sigma (i\omega_n)=-\frac{\hbar}{\omega_n}
\lim_{n\rightarrow 0}\frac{1}{n}\int d^d(\vec x-\vec
x')\int_{0}^{\beta} d(\tau-\tau')
\frac{\delta^2\overline{Z^n}}{\delta A_x(\vec x,\tau) \delta
A_x(\vec x',\tau')} e^{i\omega_n(\tau-\tau')} \eeq involves two
derivatives of the free energy with respect to the vector potential.
The resultant Kubo formula for the conductivity\cite{schon,magcon}
\beq
\sigma(i\omega_n)&=&\frac{(e^\ast m_H^2)^2}{2 h\beta\omega_n}\sum_{a,b,\omega_m}\sum_{l=0}^\infty(l+1)\left[2G_0^{ab}(l,\omega_m)G_0^{ab}(l+1,\omega_m)\right.\nonumber\\
&&\left.-G_0^{ab}(l,\omega_m)G_0^{ab}(l+1,\omega_m+\omega_n)-G_0^{ab}(l,\omega_m+\omega_n)G_0^{ab}(l+1,\omega_m)\right]
\eeq involves an explicit sum, $l$, over the Landau levels.  In the
absence of the glassy landscape (that is, $q^{ab}=0$), this
expression has been evaluated previously\cite{schon,magcon}.  The
result in the quantum disordered phase\cite{magcon,schon} is that an
insulating phase obtains regardless of the magnitde of the
dissipative dynamics $|\omega|$.  Consequently, the only terms that
might give rise to a metallic phase are those proporotional to the
Edwards-Anderson order parameter, $q$.  The leading such term gives
rise to a conductivity of the form, \beq\label{cond2D}
\sigma^{2D}(\omega=0,T\rightarrow 0)&=&\frac{(e^\ast
m_H^2)^2}{h}\sum_{l=0}^\infty (l+1)
\frac{2q\eta}{(m^2+(l+\frac{3}{2})m_H^2)^2(m^2+(l+\frac{1}{2})m_H^2)^2}\nonumber\\
&=&\frac{(e^\ast
m_H^2)^2}{h}\frac{1}{m^4}\left(\frac{2}{x}-\frac{\Psi(1,\frac{x+2}{2x})}{x^3}\right).
\eeq Where $x$ is defined as $x=\frac{{m_H}^2}{m^2}$. The summation
over $l$ give rise a digamma function defined as $\Psi(1,x)=\frac{d
\ln \Gamma(x)}{d x}$ where $\Gamma(x)$ is the Gamma function. The
large and small field limits relative to $m^2$ can be found exactly.
In the weak field regime, we find that the dc conductivity
\beq\label{condh2} \sigma^{2D}(\omega=0,T\rightarrow 0)=
\frac{4e^2}{3
h}\frac{q\eta}{\tilde{m}^4}\left(1+\frac{m_H^2}{\tilde{m}^2}\right)
\simeq \frac{4e^2}{3 h}\frac{q\eta}{{m}^4}\left[1-\frac{7}{20}x^2
\right] \quad x\equiv m_H^2/m^2\ll 1 \eeq can be written compactly
in terms of the renormalized mass scale, $\tilde{m}$. In obtaining
this expression, we worked in the correct order of limits\cite{ds}
for the dc conductivity; that is, $\omega=0$, $T\rightarrow 0$. When
the field vanishes, this conductivity goes over smoothly to Eq.
(\ref{cond1}). In the opposite limit, $m_H^2\gg m^2$, the
conductivity reduces to a simple sum \beq
\sigma^{2D}(\omega=0,T\rightarrow 0)
%=\frac{(e^\ast)^2}{h}\frac{q\eta}{m_H^4}\sum_{n=1}^\infty\frac{(n+1)}{(n+\frac{1}{2})^2(n+\frac{3}{2})^2}
=\frac{(e^\ast)^2}{h}\frac{q\eta}{m_H^4}\left[2+\frac{\pi^2}{x}
\right],\quad x \gg 1 \eeq

Consequently, we find then that the conductivty of the
Bose metal remains finite in both the weak and strong field regimes.

\subsection{Transport in 3D}

\begin{figure}
\includegraphics[scale=0.45,angle=270]{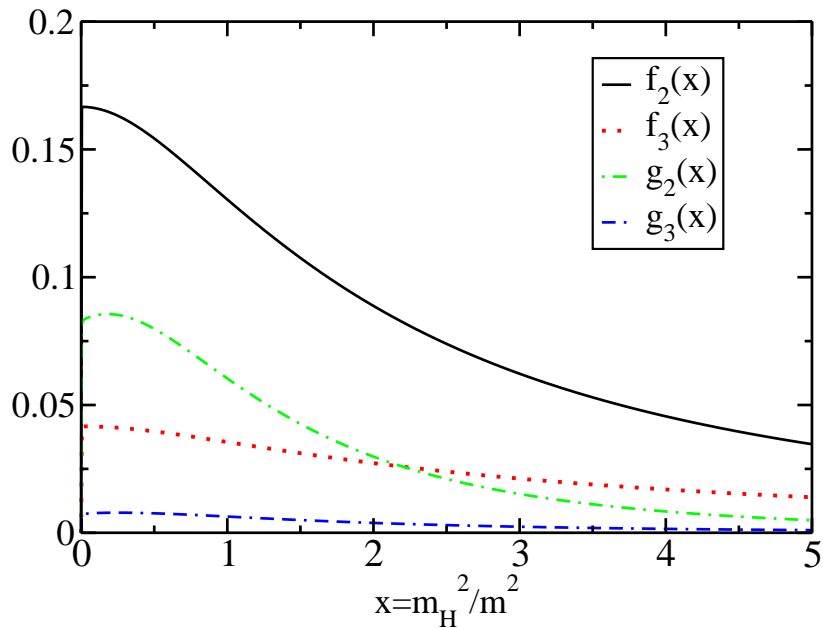}
\caption{ Conductivity of in the presence of magnetic field. In 2D, the longitudinal conductivity is given by $\sigma^{2D}=\sigma_0 f_2(x)+\sigma_1 g_2(x)$ with $\sigma_0=8e^2q\eta/hm^4$ and $\sigma_1=8e^2q^2U\eta/h\pi m^6$.
The corresponding expression in 3D is $\sigma^{3D}=m\sigma_0 f_3(x)+m\sigma_1 g_3(x)$.\label{condfig}}
\end{figure}

That the fluctuation conductivity reaches a plateau in the phase glass in the
presence of a magnetic field does not appear to have been expected previously. Before we consider the effect of the quartic term, we show that the metallic state persists in three spatial dimensions as well. The key in generalizing the formalism presented thus far to 3D is to include in the expansion for the order parameter,
\beq\label{psi3d}
\psi^a(l,x,y,z,p_y,p_z)=e^{ip_y y+ip_z z}\psi_l(x) C^a(l,p_y,p_z)
\eeq
a plane wave for the third direction.  For a vector potential of the form, ${\bf A(r)}=(0,Hx,0)$, the spatially varying part of the action involves the eigenvalue problem,
\beq
-\frac{\partial^2}{\partial x^2}\psi_l(x)+\left(\frac{e^\ast H}{\hbar}-x_0\right)^2\psi_l(x)=\tilde{E}_l(p_z)\psi_l(x)
\eeq
where
\beq
\tilde{E}=m_H^2(l+\frac12)+p_z^2.
\eeq
Except for the $p_z$-dependence, this oscillator problem is identical to the 2D analogue. In fact, expanding the free energy in terms of the states in Eq. (\ref{psi3d}) and computing the derivatives with respect to the vector potential
in the Kubo formula reveals that the primary difference in the longitudinal conductivity
\beq
\sigma^{3D}(i\omega_n)&=&\frac{(e^\ast m_H^2)^2}{2 h\beta\omega_n}\sum_{a,b,p_z,\omega_m}\sum_{l=0}^\infty(l+1)\left[2G_0^{ab}(l,\omega_m,p_z)G_0^{ab}(l+1,\omega_m,p_z)\right.\nonumber\\
&&\left.-G_0^{ab}(l,\omega_m,p_z)G_0^{ab}(l+1,\omega_m+\omega_n,p_z)-G_0^{ab}(l,\omega_m+\omega_n,p_z)G_0^{ab}(l+1,\omega_m,p_z)\right]
\eeq
is the presence of a sum over $p_z$. Because the $p_z$ dependence in
the bare Green functrion
\beq
G_0 (l,p_z,\omega_n)=(l m_H^2+\omega_n^2+\eta|\omega|+\tilde{m}^2+p_z^2)^{-1}
\eeq
amounts to a shift in $m^2$, the final expression for the conductivity in
$D=3$
\beq
\sigma^{3D}(\omega=0,T\rightarrow 0)=\frac{(e^\ast m_H^2)^2}{h}\sum_{p_z}\sum_{l=0}^\infty (l+1) \frac{2q\eta}{(m^2++p_z^2+(l+3/2)m_H^2)^2(m^2+p_z^2+(l+1/2)m_H^2)^2}.
\eeq
is easily obtained from Eq. (\ref{cond2D}).
Let us define
\beq
A_n=m^2+\left(n+\frac12\right)m_H^2.
\eeq
Performing the integral over $p_z$ using residues reduces the conductivity
\beq\label{sigma3d}
\sigma(\omega=0,T\rightarrow 0)=\frac{\pi (e^\ast)^2}{2\hbar}m_H^4\sum_{n=0}^\infty (n+1)\left[\frac{1}{A_n^{3/2}A_{n+1}^{3/2}(A_n^{1/2}+A_{n+1}^{1/2})}+\frac{1}{A_nA_{n+1}(A_n^{1/2}+A_{n+1}^{1/2})^3}\right]
\eeq
to a single sum over the Landau index.  As in the 2D case, this expression can be evaluated analytically only in the small and large field limits.  Defining
$x=m_H^2/m^2$, we find that the strong-field limit is given by
\beq
\sigma^{3D}= \frac{(e^\ast)^2}{\hbar}\frac{q\eta\pi}{m^3}\left[1+\frac{1}{ x^{3/2}}\left(\zeta(5/2)-\frac{\zeta(7/2)}{2}\right)\right]\quad x\equiv m_H^2/m^2\gg 1,
\eeq
where we have made the approximation,
\beq
\sum_1^\infty\frac{1}{(n+1/2)^m(n+3/2)^{1/2}}\approx \zeta\left(\frac{2m+1}{2}\right).
\eeq
In 3D, the leading asymptotic inverse correlation length dependence scales as
$m^{-3}$ as opposed to the $m^{-4}$ dependence obtained in 2D.  In the opposite limit,
\beq
\sigma^{3D}=\frac{(e^\ast)^2}{\hbar}\frac{q\eta\pi}{m^3}\left[\frac16+\frac18
 x\right]\quad x\ll 1.
\eeq
For the sake of comparison, we have evaluated the conductivities in 2D and 3D numerically and plotted them in Fig. (\ref{condfig}) as a function of x.    As is observed experimentally, both are decreasing functions of the magnetic field.

\subsection{Role of interactions:}

\begin{figure}
\includegraphics[scale=0.45]{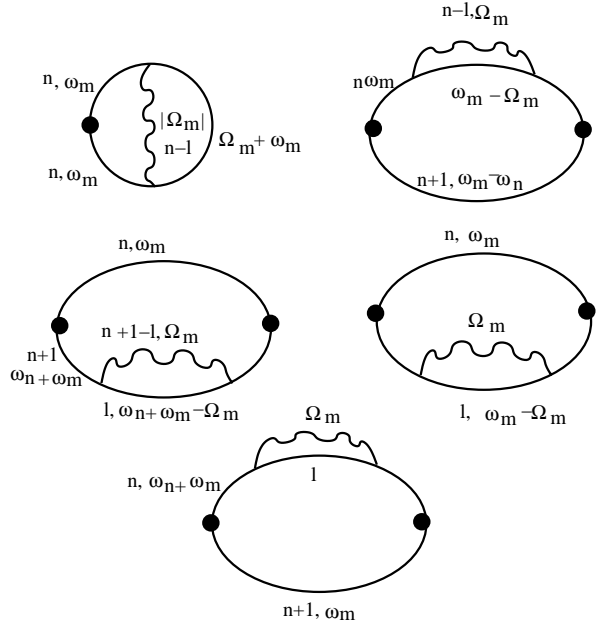}
\caption{ One-loop diagrams that determine the leading interaction corrections to the conductivity. The squiggly line represents the quartic interaction.  The matsubara frequencies and Landau level indices are indicated. These determine the
 conductivity correction in Eq. (\ref{intsum}).\label{vertex}}
\end{figure}

Thus far, we have worked at the Gaussian level in which interactions
are strictly ignored.  At the tree level, a dynamical exponent of
$z=2$ renders the quartic interaction $U$ marginally irrelevant.
However, considering the last term in Eq. (\ref{fgauss}) on equal
footing with $U$ in the one-loop renormalization group scheme, we
reach the conclusion that the RG equations flow to strong coupling.
The relevance of $q$ at all dimensions manifests itself also by the
increasing singularity of relevant contributions from higher order
diagrams in the perturbation series in $U$.  Consequently, we
consider the role of interactions. Consider first the linear $U$
correction which is given by the diagrams in Fig. (\ref{vertex}). We
use the definition of $A_n$ to simplify the notation in the
evaluation of the diagrams.
\noindent
In two dimensions, the sum of these diagrams is given by
\beq\label{intsum}
\sigma_1^{(1)}&=&\frac{(e^\ast)^2}{2h}\frac{q^2Um^2_H}{\beta\omega_n}\left\{2\sum_n(n+1)(A_n)^{-3}A_{n+1}^{-1}+A_{n+1}^{-3}A_{n}^{-1}+2A_{n}^{-2}A_{n+1}^{-2})\right.\nonumber\\
&&\left.-\sum_n(n+1)m_H^2\left(2A_{n+1}^{-2}G^2_0(n,\omega_n)+4A_{n}^{-3}G_0(n+1,\omega_n)\right)\right.\nonumber\\
&&\left.-\sum_n
(n+1)m_H^2\left(2A_{n}^{-2}G_0^2(n+1,\omega_n)+4A_{n+1}^{-3}G_0(n,\omega_n)\right)\right\}\sum_lA_{l}^{-2},
\eeq which simplifies in the zero-frequency limit to
\beq
\sigma_1^{(1)}&=&\frac{32e^2}{h} \frac{q^2U\eta}{4\pi
m^6}\sum_{n,l}\frac{(n+1)x^3}{A_n^3A_{n+1}^2A_l^2}.
\eeq
Two
relevant features are that the leading $U$ correction has a
non-divergent contribution to the conductivity in the $T\rightarrow
0$ limit and the dependence on $m$ is more singular than in the
$U=0$ case.  This points to a scaling function of the form $\sigma
\approx (e^2/\hbar)(\eta q/m^4)\Phi\left( q /m^2 \right)
f(U,\omega_H)$ where $\Phi(y) \sim y^p$ for large $y$, which yields
the critical behavior $\sigma \sim m^{-x}$ with $x = 4+2p$ and
$f(U,\omega_H)$ is a general some function of the interactions and
the magnetic field.  The value of the exponent $p$ cannot be inferred
at any finite order in perturbation theory.

The analogous expression can be derived for the three-dimensional case.  The only complication here is the integration over $p_z$.  The result,
\beq
\sigma_{\rm 3D}(\omega_n=0)&=&\frac{4 e^2}{h}\frac{q^2\eta U}{ m^5}\sum_{n,l}
\left[\frac{3\pi}{8}\frac{n+1}{A_n^{5/2}}+\frac{\pi}{2}\frac{(n+1)(A_nA_{n+1}^{1/2}-2A_n^{1/2}A_{n+1})}{A_n^{3/2}A_{n+1}^{3/2}(A_n^{1/2}+A_{n+1}^{1/2})^2}\right.\nonumber\\
&&\left.-\frac{\pi}{2}\frac{(A_n+\omega_H)(n+1)}{A_n^{3/2}A_{n+1}^{3/2}(A_n^{1/2}+A_{n+1}^{1/2})}\right]\frac{1}{A_l^{3/2}},
\eeq expressed in terms of the function $A_n$ is once again finite
as $T\rightarrow 0$.  Hence, the bosonic conductivity is robust to
interactions (at least in leading order) in both $D=2$ and $D=3$ in
the presence of a magnetic field.

\begin{figure}
\includegraphics[scale=0.45,angle=270]{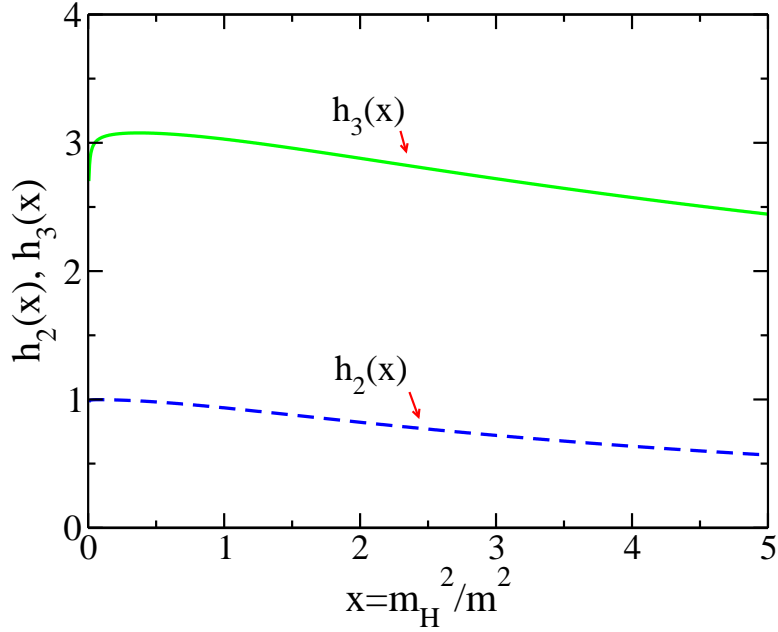}
\caption{Disorder contribution to the renormalized gap.  In 2D,
$\delta m^2_{2D}=(qU/4\pi m^4) h_2(x)$ whereas in 3D, $\delta m^2_{3D}=(qU/4\pi m^3) h_3(x)$ . \label{pgmass}}
\end{figure}

Of course, the mass, $m$, is also renormalized by the interactions.
To understand its dependence on the interactions at large $N$. The
self-consistent $1/N$ expansion for the correction to the mass gap,
\beq \frac{\delta m^2}{U}=\frac{T m^2_H}{4\pi}\sum_{\omega_m,n}
G_{0}(\omega_m,n)+\beta q\delta_{\omega_m,0} G_{0}(\omega_m,n)^2
\eeq contains two terms.  The term independent of the glass
environment has been calculated previously\cite{dp2}. The second
term, written succinctly as, \beq \frac{\delta
m^2_{2D}}{U}=\frac{\omega_H T\beta q}{4\pi
m^4}\sum_n\frac{1}{(1+(n+1/2)x)^2} \eeq is explicitly independent of
temperature and provides hence only an added magnetic field
dependence to the gap equation.  Here $x=m^2_H/m^2$.  In the 3D
case, the integral over $p_z$ leads to slightly different power in
the denominator, \beq \frac{\delta
m^2_{3D}}{U}=\frac{q}{8m}\sum_{n=0}^\infty\frac{x}{(1+(n+1/2)x)^{3/2}},
\eeq with, once again, no added temperature dependence.  Hence, the
interaction terms that couple to the glassy degrees of freedom lead
to an innocuous renormalization of the mass as depicted in Fig.
(\ref{pgmass}).

\subsection{Absence of phase stiffness}

In the strict sense, a physical system possesses a non-zero phase
rigidity if upon a non-uniform rotation of the phase, the free energy
increase is of the form,
\beq\label{eq1} \Delta
F=\frac{\rho_s}{2}\int d^2r|\nabla\theta|^2,
\eeq where $\rho_s$
is
the spin or superfluid stiffness and $\theta$ is the collective
phase variable.  For the problem at hand, it suffices to expand the
gradient part of the free energy for the spin-glass part
in terms of a local part given by Eq. (\ref{qm}) and a
spatially varying part we refer to as
$\tilde{Q}_{\mu\nu}^{ab}(k,\omega_1,\omega_2)$.

%=========================================================================
% some new input on phase stiffness
%=========================================================================
Firstly, let us consider the contribution of the spatial independent
part of $Q$. It has been shown recently\cite{fl} that a $Q^4$th term which is responsible for replica symmetry breaking in the spin glass action changes the mean-field solution for the excitation spectrum, $D(\omega)$. In this case,
the excitation spectrum,
\beq\label{newD}
D(\omega)=c q_{EA}^2-|\omega|/\kappa
\eeq
remains gapped in the spin glass phase.  Here
 $c$ is a constant and $q_{EA}$ is the diagonal part of
$q^{ab}$. In the absence of the $Q^4$th term, the stiffness has been shown to vanish\cite{qstiff}.  However, recently, Ikeda\cite{ikeda2} has disputed this claim and has proposed that the constant term in Eq. (\ref{newD}) gives
leads to a non-zero stiffness.  We show that this claim is false.
 Following our earlier calculation\cite{qstiff} (which Ikeda\cite{ikeda} also follows), we find that the relevant
contribution to the spin-glass stiffness containing the $D(\omega)$ term reduces to
\begin{equation}
\sigma_{PS}(i\omega_n)=\frac{16e^2}{\hbar
\omega_n}\left[T\sum_{\omega_m} D(\omega_m)\left(
D(\omega_m)-D(\omega_m+\omega_n)\right)-q_{EA}\left(D(\omega_n)-D(0)\right)\right]
\end{equation}
in the presence of the constant term. 
As we have shown previously\cite{qstiff}, $\sigma_{\rm PS}$ can be evluated
by analytically continuing the first term using the definition of $D(\omega)$.  The result\cite{qstiff} is non-critical and metallic even in the presence of the constant contribution to $D(\omega)$.  This state of affairs arises because as long as difference of $D$'s always appear, any constant that is added to $D(0)$ cannot yield a divergent contribution to the spin-glass conductivity, contrary to the claims of Ikeda\cite{ikeda}. 

%=========================================================================
 The contribution of spatially dependent part of $Q$ can
be extracted from the propagator for the spatially varying part of
the action. Note that from Eq. (\ref{qlin}), $Q^{ab}_-$ is
determined by the longitudinal components of the spin-glass
order parameter only. Further, $Q^{ab}_-$ does not couple to the
magnetic field. The transvere components enter $Q^{ab}_+$ only and
couple directly to the field.  Hence, it is only the {\bf spatial} fluctuations of
$Q^{ab}_+$ that are of interest here.  The associated  propagator
for $\tilde{Q}_+^{ab}(k,\omega_1,\omega_2)$ at the Gaussian level is
simply
\beq g_+^{ab}(n,\omega_1,\omega_2)=\frac{1}{(n+\frac12)m_H
+|\omega_1|+|\omega_2|}.
\eeq
Because $\omega$ is a Matsubara
frequency, the propagator describes the motion of mode whose energy
dispersion scales as $i(n+1/2)m_H^2$.  Such a mode is purely imaginary and
hence does not possess a stiffness.  For $Q^{ab}_-$, $(n+1/2)m_H^2$ is
replaced by $k^2$ and diffusive transport ensues.  Hence, neither the transverse nor the longitudinal components possess a stiffness.  Consequently, the phase
glass is not a superconductor and only the fluctuation conductivity
computed previously contributes to the dc transport. A corrolary of this work is that the vortex glass, in which the disorder in the $J_{ij}$'s (caused by a random vector potential) is bounded between $[-1, 1]$ does not have a stiffness and hence is not a superconductor.

\begin{figure}
\includegraphics[scale=0.65,angle=0]{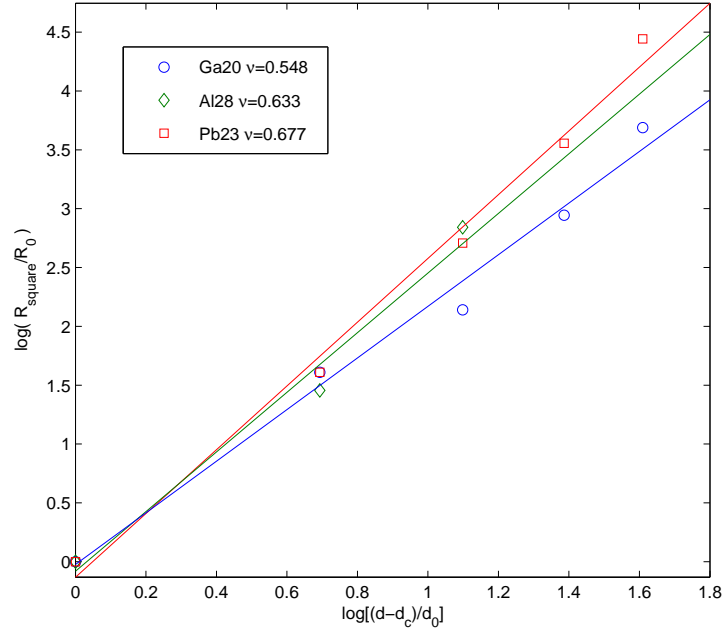}
\caption{Critical scaling of the turn-on of the resistivity near the superconductor-metal transition on the metallic side as a functin of the sample thickness.  $d_c$ represents the critical thickness needed to destroy superconductivity. Based on Eq.
(\ref{cond1}), we are able to extract a value of the correlation length exponent, $\nu$.  The fits establish a value of $\nu$ roughly between $[1/2,2/3]$.  The data were extracted from Ref. (1). \label{nudata}}
\end{figure}

\subsection{Comparison with Experiment}

\begin{figure}
\includegraphics[scale=0.65,angle=0]{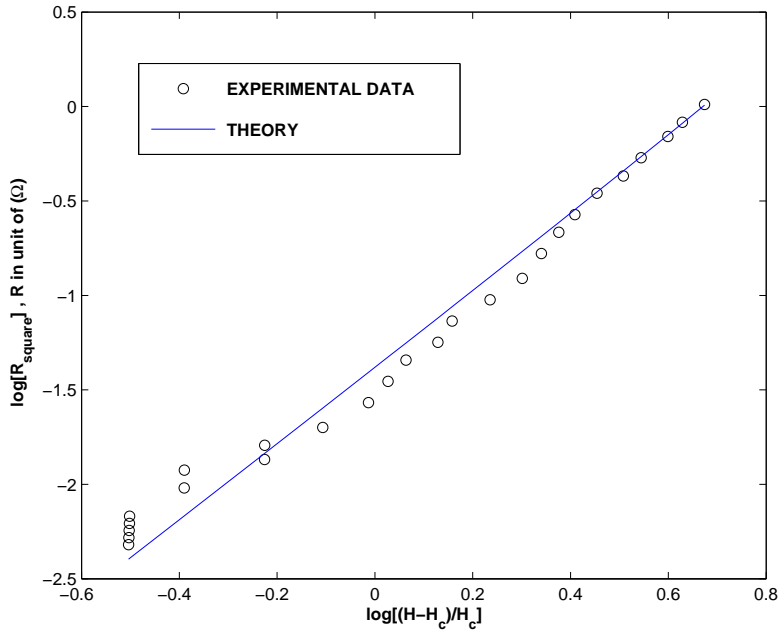}
\caption{The turn-on of the resistivity as a function of magnetic field for MoGe near the superconductor-metal transition on the metallic side. The theory (slid line) curve is determined by Eq. (\ref{lowhexp}).  The best-fit curve, $70\Omega (H-H_c)^{2.417}$ is in excellent agreement with the theory and with the experimental result of $R_c(H-H_c)^\mu$, $1<\mu<3$.  The data were taken from N. Mason's thesis\cite{masonthesis}.\label{mkdata}}
\end{figure}

Our theory makes precise claims as to how the resistivity turns on
in both the disordered and magnetic-field tuned transitions.  Hence,
given the plethora of experimental data, it behooves us to put our
predictions to the test.  Consider first the disorder-tuned
transition.  In this case, the disorder resistivity turns on as the
fourth power of the mass according to Eq. (\ref{cond1}).  As
$m^2\propto \xi^{-z}$ and $\xi\propto (d-d_c)^{-\nu}$, we find that
the conductivity turns on as $(d-d_c)^{2z\nu}$. For the problem at
hand, the mean-field exponents are given by $z=2$ and $\nu=1/2$.
Any attempt to go beyond mean-field will result in an uncontrolled
expansion as a result of the coupling to the $Q^{ab}$ degrees of
freedom\cite{rsy}.  If a controlled expansion were possible, $\nu$
would be renormalized not $z$.  Hence, to understand how close the
transition is to mean-field, we treat $\nu$ as a fitting parameter.
Consequently, we used the expression $\rho(T=0)\propto
(d-d_c)^{4\nu}$ to fit the data on three different samples. The
results shown in Fig. (\ref{nudata}) point to a value of $\nu$
ranging between $[1/2,2/3]$.  To corroborate this prediction, we
turn to the magnetic field tuned transition. Mason and
Kapitulnik\cite{masonthesis} find that the resistance per square
turns on as \beq\label{masonfit} R\propto R_c(H-H_c)^{\mu}\quad
1<\mu<3. \eeq   Since we are interested in the region close to the lower
critical field, it suffices to investigate the critical properties
of the conductivity using the weak-field expression, Eq.
(\ref{condh2}).  A rough scaling of the conductivity can be obtained
by noting that in the weak-field expression, $m^2$ is replaced by
$m^2+m_H^2=\tilde{m}^2$. If we replace $\tilde{m}^2$ by $h^{-z\nu}$,
with $h=(H-H_c)/H_c$, we find that $\rho\approx h^{4\nu}$.   To fit
Eq. (\ref{masonfit}), we require that $4\nu\approx 2.6$ or
alternatively, $\nu=.63$, a value not inconsistent with the
disorder-tuned transition, though it is unclear the precise magnitude of
the error bars in this fitting procedure.  Consequently, we resort
to a more rigorous treatmet.   The transition to the metal state
occurs when $m^2+m_H^2=0$.  For $m^2<0$, we can define
$x=m^2_H/|m^2|$. Then the Eq. (\ref{condh2}) can be written as \beq
\sigma^{2D}(\omega=0,T\rightarrow 0)=\frac{(e^\ast m_H^2)^2}{h
m^8}\sum_{l=0}^\infty
\frac{2q\eta(l+1)}{(-1+(l+\frac{3}{2})x)^2(-1+(l+\frac{1}{2})x)^2}=\frac{(e^\ast
m_H^2)^2}{h}\frac{1}{m^4}\left(\frac{2}{x}+2\frac{\Psi(1,\frac{x-2}{2x})}{x^3}\right)
\eeq The first singularity appears at $x=2$ which is the
superconductor-metal transition point and it yields a critical
magnetic field $H_c=\frac{m^2 \hbar}{e^*}$. Expanding the
conductivity around $x=2$, we obtain for $h\rightarrow 0$
\beq\label{lowhexp}
\rho_{xx}=\frac{1}{\sigma_{xx}}=2\pi\frac{\hbar}{(e^*)^2}\frac{m^4}{q\eta}\left(\frac{h^2+h^3}{2}\right)
\eeq Hence, the leading term is $h^2$ which is consistent with the
turn-on of the resistivity in the disorder-tuned transition.  A
comparison of this data and a best fit plot to the magnetic-field
tuned data is shown in Fig. (\ref{mkdata}). The consilience of these
results unifies the disorder and magnetic-field tuned transitions.
Both have the same cause and their turn-on exponents resistivity as
a function of the disorder and magnetic field are equivalent.  Our
predicition that $\nu\approx 1/2$ can certainly be tested further
experimentally.  In comparing with experiment, it is certainly best to know the width of the critical region.  However, since heat capacity measurements are not available, we can only obtain a lower bound using the standard Ginzburg criterion.  In our work, we calculated the conductivity in the zero temperature limin terms of the inverse correlation length, $m^2$ as, which to leading order in
the field is given by
\beq
        {\widetilde{m}}^2(T=0)=m^2+\frac{e^* H}{\hbar c}=\alpha(-T_c)+\frac{e^* H}{\hbar c}=\frac{e^* H_c}{\hbar c} \frac{(H-H_c)}{H_c}=\frac{e^* H_c}{\hbar c} h
\eeq
Substituting this expression into the Ginzburg criterion and assuming that the specific heat term is of order unity, we find that $h\equiv (H-H_c)/H_c\approx 10^{-6}$ sets the lower bound for the width of the critical region.
\section {Global Summary}

%==================================================================================
% Table for the conductivity of all the bosonic model
%==================================================================================

All the bosonic models considered here are obtained from
an action of the form,
\beq
S_{\psi}&=&\int d^dx\left\{ d\tau \sum_a\Psi_\mu^a(x,\tau)
\left[ m^2-\left(\nabla-\frac{i e^\ast}{c\hbar}A(x,\tau)\right)^2\right]\Psi_\mu^a(x,\tau)\right\}\nonumber\\
&&+\int d^dx\int
d\tau\sum_a\frac{\partial}{\partial\tau_1}\Psi_\mu^a(x,\tau_1)
\frac{\partial}{\partial\tau_2}\Psi_\mu^a(x,\tau_2)|_{\tau_1=\tau_2=\tau}
+ U \int
d\tau\sum_a\left[\Psi_\mu^a(x,\tau)\Psi_\mu^a(x,\tau)\right]^2\nonumber\\
&& +\eta \int d^d x\int d\tau_1 d\tau_2\sum_{a}\Psi_\mu^a(x,\tau_1)
|\frac{\partial}{\partial \tau_2}|
\Psi^a_\nu(x,\tau_2)|_{\tau_1=\tau_2=\tau}+\beta \int d^d x\int
d\tau \sum_{a,b} q^{a b} \Psi_\mu^a(x,\tau) [\Psi^b_\nu(x,\tau)]^* .
\eeq

The parameters which determine the conductivity are the
dissipation $\eta$, the Edwards-Anderson order parameter $q^{a b}$, the quartic
 interaction $U$, temperature
$\beta=1/K_B T$ and magnetic field $B$. In the following, we provide a
brief table which catalogues all the distinct cases that arise and the
dependence of the conductivity on the above factors.  The table exhausts all
known Bose models in which the conductivity has or can be calculated.
\begin{table*}[bottom]
%[bottom]
\caption{\label{BoseConTable} Conductivity for bosonic models as a
function of system paraments.}
\begin{ruledtabular}
\begin{tabular}{ l l l l l l }
%l@{\qquad\qquad}l@{\,}l@{\,}l@{\,}l@{\quad}l}
%Symmetry & \multicolumn{5}{c}{Expression in terms of
%left(L)- and right(R)-moving components} \\
%\hline
%\hline
\multicolumn{3}{c }{Distinct Cases } & $\eta=0\ \  q=0$ &
$\eta\neq
0\ \ q=0$ & $\eta\neq 0\ \  q\neq 0$ \\
\hline
%==========================================================================
$B=0$ & $T=0$ &$U=0$ & $0$ \textbf{I}  & $0$ \textbf{I} &
$\frac{e^2}{h}\frac{4q\eta}{3 m^4} $ \textbf{M}  \\
%\cline{3-5}
%---------------------------------------------------------------------------
$B=0$ &$T=0$ &  $U\neq 0$& $0$ \textbf{I}  & $0$ \textbf{I}(QD)
$\infty$ \textbf{SC}(RC) \cite{dp}&$\frac{e^2}{h}\frac{4q\eta}{3
m^4}(1+\frac{qU}{\pi m^2}) $ \textbf{M}
 \\
 %\cline{2-6}
%============================================================================
 $B=0$ & $T\neq 0$ &$U=0$ & $\infty$ \textbf{PC}  \cite{avan}  & $\frac{4e^2}{\pi
h}[\frac{\pi T}{\eta}e^{-m/T}+(\frac{\pi\eta T}{3 m^2})^3]\ (T\ll
m)$  & Box left+Box above\footnotemark[2]\\
& & & &$\frac{2e^2 T}{h \eta}\ln(\frac{T}{m})(\eta\ll m;m\ll T)$ & \\
& & & &$\frac{e^2 \eta T}{h m^2}(\eta\gg m;m\ll T)$  \cite{dp} & \\
%\cline{3-6}
%--------------------------------------------------------------------------------------
 $B=0$ &$T\neq 0$ &$U\neq 0$ & $\infty$ \textbf{SC}  \cite{scalapino} & $U$ renormalize $m$ \cite{dp}\footnotemark[1]
   & $U$ and $q$ renormalize $m$\footnotemark[1] \\
\hline
%==============================================================================
$B\neq 0$ & $T=0$ &$U=0$ & $0$ \textbf{I} & $0$ \textbf{I} &
$\frac{e^2}{h}\frac{4q\eta}{3\tilde{m}^4}
(1+\frac{2 m_H^2}{\tilde{m}^2})$ \textbf{M} \\
%\cline{3-5}
%-------------------------------------------------------------------------------
$B\neq 0$  &$T=0$ & $U\neq 0$ & 0 \textbf{I} & 0 \textbf{I} or
$\infty$ \textbf{SC} & $\frac{e^2}{h}\frac{4q\eta}{3\tilde{m}^4}
(1+\frac{2 m_H^2}{\tilde{m}^2}+\frac{qU}{\pi \tilde{m}^2}f(x))$ \textbf{M}\\
%\cline{2-6}
%===============================================================================

$B\neq 0$  & $T\neq 0$ & $U=0$ & $\infty$ \textbf{PC} &
$\frac{4e^2}{\pi
h}[\frac{\pi T}{\eta}e^{-\tilde{m}/T}+(\frac{\pi\eta T}{3 \tilde{m}^2})^3] $   & \\
& & & & $(m^2_H\ll \tilde{m}\eta ;T\ll \tilde{m})$   & Box left+Box above\footnotemark[2]\\
& & & &$\frac{4e^2}{
h\eta }[\frac{8\eta T\tilde{m}^2}{m_H^4}e^{-\tilde{m}/T}+(\frac{\pi\eta T}{3 \tilde{m}^2})^3]$ &  \\
& & & &$(\eta \tilde{m} \ll m_H^2\ll \tilde{m} T;T\ll \tilde{m} )$
&\\

& & & & $\frac{2e^2 T}{
h\eta }[\ln(\frac{T}{\tilde{m}})+\frac{11 m_H^2}{32 \tilde{m}^2}]$  &  \\
& & & & $(\tilde{m} \ll T;\eta\ll \tilde{m},m_H)$ &\\
& & & & $\frac{e^2\eta T}{
h\tilde{m}^2}[1+\frac{m_H^2}{2 \tilde{m}^2}]$ &\\
& & & & $(\tilde{m} \ll \eta \ll m_H\ll T)$ \cite{denisu} &  \\
%\cline{3-6}
%------------------------------------------------------------------------------------
$B\neq 0$ & $T\neq 0$&  $U\neq 0$ & $\infty$ \textbf{SC} & $U$ and
$B$ renormalize $m$\footnotemark[1]
& $U$,$B$ and $q$ renormalize $m$\footnotemark[1] \\
%\hline

\end{tabular}
\end{ruledtabular}
\footnotetext[1]{$U$,$B$ and $q$ will renormalize $m$ and change the
dependence of temperature $T$ thus change the behavior of the
conductivity indirectly. See section:Role of interaction.}
\footnotetext[2]{The conductivity in this case is equal to $\sigma$
($\eta\neq0$, $q=0$,$T\neq 0$,$U=0$) plus $\sigma$($\eta\neq0$,
$q\neq 0$,$T=0$,$U=0$).} \footnotetext[3]{\textbf{PC}, \textbf{SC},
\textbf{I}, \textbf{M},QD and RC represent perfect conductor,
superconductor, insulator, metal, quantum disorder regime and
renormalized classical regime respectively.}
\end{table*}

%===================================================================================
% End of Table
%===================================================================================

\section{Final Remarks}

We have shown here that the bosonic conductivity in the phase glass is robust to both interactions and a non-zero magnetic field. Further, we find complete agreement with between theory and experiment for the disorder-tuned and the magnetic-field tuned turn-on of the resistivity in 2D films. In general, the resistivity turns on as
\beq
\rho(T=0)\propto \rho_c(g-g_c)^{2z\nu}
\eeq
where $g$ is either disorder or magnetic field. The significance that a single expression with universal exponents describes both the disorder-tuned data as well as the field-tuned data is two-fold: 1) both the disorder and magnetic-field transitions share a common origin and 2) both are driven by superconducting fluctuations, that is, both phases are bosonic rather than fermionic in character.  In fact, the power-law turn-on of the resistivity places severe restrictions any permissible theory of the Bose metal.  Further, identical results hold in D=2 as well as in D=3.  Hence, the phase glass is a candidate to explain the experimental findings of the leveling of the resistivity as $T\rightarrow 0$ in thin films that nominally display a superconductor-insulator transition.
The fact that the metallic phase persists in D=3 is particularly revealing in light of the recent experiments which find that in the putatative vortex glass phase, the $I-V$ characteristics remain linear as opposed to the non-linear behaviour anticipated for a true superconducting phase.  As it turns out, no explicit calculation of the conductivity has ever been performed for the vortex glass.  The only difference between the vortex glass and the phase glass considered here is that in the former, the disorder in the $J_{ij}$'s is bounded in the interval $[-1,1]$.  This is not a substantive difference. Hence, the model considered here contains the complexity of the vortex glass and hence should suffice to describe this model.  Further, as we have shown above and elsewhere\cite{qstiff}, coupling the electromagnetic gauge directly to the spin glass order parameter does not proves conclusively that the stiffness vanishes.  In light of this result and the results derived here, we conclude that the vortex glass is in fact a metal.  Ultimately, if our proposal is correct, the metallic phase in the thin films has been seen before.  Namely, it is the vortex glass.

\section{Appendix A:}
%=================================================================================================================
% This part is the equations refered in the appendix. Please pay attension to the label. 
%=================================================================================================================
The propagator, Eq. (\ref{prop}), describes bosonic transport for the free energy given by Eq. (\ref{fgauss}):
\beq\
{\cal F_{\rm gauss}}&=&\sum_{a,\vec k,\omega_n}
(k^2+\omega_n^2+\eta |\omega_n|+m^2)
|\psi^a(\vec k,\omega_n)|^2 \nonumber\\
&&-\beta q\sum_{a,b,\vec k,\omega_n} \delta_{\omega_n,0}
\psi^a(\vec k,\omega_n)[\psi^b(\vec k,\omega_n)]^{\ast},
\eeq
Consequently, in replica space, the inverse propagator is
\begin{equation}\label{ginv}
G^{(0)}_{ab}(\vec k,\omega_n)={\left( m^2+\vec k^2+\omega^2+\eta |\omega|-\beta q^{ab}\delta_{\omega,0} \right)}^{-1}_{ab} .
\end{equation}
That is,
\begin{equation}
         \sum_b G^{(0)}_{ab}(\vec k,\omega_n)\left( m^2+\vec k^2+\omega^2+\eta |\omega|-\beta q^{bc}\delta_{\omega,0} \right)=\textbf{1}_{ac}
\end{equation}

To invert the matrix, we set 
\beq
G^{0}_{ab}(\vec k,\omega_n)=A(\vec k,\omega_n)\delta_{ab}+B(\vec k,\omega_n)\beta q^{ab}
\eeq
 and 
\beq
C(\vec k,\omega_n)= (m^2+\vec k^2+\omega^2+\eta |\omega|)
\eeq
and then substitute into Eq. (\ref{ginv}) to obtain
\beq\label{invt}
\textbf{1}_{ac} &=&\sum_b       \left(A(\vec k,\omega_n)\delta_{ab}+B(\vec k,\omega_n)\beta q^{ab}\right)
\left(C(\vec k,\omega_n)\delta_{bc}-\beta q^{bc}\delta_{\omega,0} \right)\nonumber\\
&=&A(\vec k,\omega_n)C(\vec k,\omega_n)\delta_{ac}+\left(B(\vec k,\omega_n)C(\vec k,\omega_n)-A(\vec k,\omega_n)\delta_{\omega,0}\right)\beta q^{ac}+B(\vec k,\omega_n)\delta_{\omega,0}\sum_b q^{ab}q^{bc}.
\end{eqnarray}
The last term is a product of Parisi matrices which is parameterized by the function $\tilde{q}$ and $q(x)$ $(0<x<1)$, where $\tilde{q}$ and
 $q(x)$ are the diagonal and off-diagonal elements of $q$ in the continous representation. The sum over the replica index $b$ in the last term in Eq. (\ref{invt}) will be proportional to $n$ and hence vanish in the $n\rightarrow 0$ limit.  Consequently, we simplify Eq. (\ref{invt}) to obtain
\begin{eqnarray}
        A(\vec k,\omega_n)C(\vec k,\omega_n)=1 \nonumber\\
        B(\vec k,\omega_n)C(\vec k,\omega_n)-A(\vec k,\omega_n)\delta_{\omega,0}=0 .
\end{eqnarray}
The solution to these equations is
\begin{eqnarray}
A(\vec k,\omega_n)=\frac{1}{C(\vec k,\omega_n)}=\frac{1}{m^2+\vec k^2+\omega^2+\eta |\omega|}=G_0 (\vec k,\omega_n)\nonumber\\
B(\vec k,\omega_n)=\frac{A(\vec k,\omega_n)\delta_{\omega,0}}{C(\vec k,\omega_n)}=\frac{\delta_{\omega,0}}{(m^2+\vec k^2)^2}= G_0^2(\vec k,\omega_n)\delta_{\omega_n,0}.
\end{eqnarray}
Eq.(\ref{prop}) is then derived.

\section{Appendix B:}

The point of this Appendix is to show that even if an Ohmic dissipation
term is put in by hand as done recently\cite{ikeda}, nothing of a substantive nature changes.  Hence, the claim that $ D\propto |\omega|^{1/2}$ leads to a superconducting state instead of a metallic one is false. To avoid any confusion,
we redo the cumulant expansion. 

The effective action for our problem is given by 
\beq\label{act}
 S_{\rm  eff}&=&\int_0^\beta
d\tau\sum_{i,a}\frac{1}{4E_C}\left(
\frac{\partial\theta_i^a}{\partial\tau}\right)^2\nonumber\\
&&+\frac{J^2}{2}\sum_{\rm a,b}\sum_{\langle ij\rangle}
\int_0^\beta\int_0^\beta d\tau d\tau'\frac{1}{4}\sum_{\alpha=+1,-1}
\exp\left\{i\left[\theta_i^{a}(\tau)-\alpha\theta_i^b(\tau')-
\left(\theta_j^{a}(\tau)-\alpha\theta_j^b(\tau')\right)-\left(A_{ij}(\tau)-
\alpha A_{ij}(\tau')\right)\right]\right\}\nonumber\\
&& +c.c.\nonumber\\
&&-\frac{J_0}{2}\sum_\alpha\sum_{ij}\int_0^\beta d\tau
\exp\left(\theta^a_i-\theta_j^b(\tau)-A_{ij}\right)+c.c.
\eeq
Using the
Hubbard-Stratanovich transformation and introducing the order
parameters $\Psi^a_\mu(k,\tau)$ and $Q^{ab}_{\mu\nu}(k,\tau_1,\tau_2)$, we have the effective free energy\cite{dp2},
\begin{eqnarray}
  F_{\rm eff}[\Psi,Q]&=& \sum_{a,k}\int_0^{\beta}d\tau \Psi_{\mu}^a(k,\tau)[\Psi_{\mu}^a(k,\tau)]^*+
  \sum_{a,b,k,k'}\int_0^{\beta}d\tau\int_0^{\beta}d\tau' Q_{\mu\nu}^{ab}(k,\tau,\tau')[Q_{\mu\nu}^{ab}(k,\tau,\tau')]^*\nonumber\\
&-&\ln \left[ \left\langle \hat{T} \exp\left(2\int^{\beta}_{0}d\tau \sum_{a,k}\sqrt{J_0(k)}\Psi_{\mu}^a (k,\tau)S_{\mu}^a (-k,\tau)\right.\right.\right.\nonumber\\
 & + & \left. \left. \left. \sum_{a,b,k,k'}\int^{\beta}_{0} d\tau \int^{\beta}_{0} d{\tau}' \sqrt{2 J(k) J(k')}Q_{\mu \nu}^{a b} (k,k',\tau,{\tau}') S_{\mu}^a (k,\tau)S_{\nu}^b (k',{\tau}')+ F_{dis} \right) \right\rangle_0  \right]\nonumber\\
 F_{dis} &=& \int^{\beta}_{0} d\tau \int^{\beta}_{0} d\tau' \frac{\eta}{\pi(\tau-{\tau}')^2} \left( 1-\sum_{a,\mu,k}S_{\mu}^{a} (k,\tau)S_{\mu}^a (-k,{\tau}')\right)
\end{eqnarray}
where $J(k)=J\sum_{R}\cos(k R)$,$J_0(k)=J_0\sum_{R}\cos(k R)$ ($R$ is the displacement vector of nearest neighbour) and
\begin{eqnarray}
        \left\langle A\right\rangle_0=\frac{1}{Z_0^n}Tr(e^{-\beta H_0^{eff}}A)
\end{eqnarray}
Here, $Z_0=Tr[\exp(-\beta H_0)]$. The Hamiltonian $H_0$ and $H_0^{eff}$ are defined as,
\begin{eqnarray}
        H_0=-\frac{1}{4E_c}\sum_i\left( \frac{\partial}{\partial\theta_i}\right)^2,\ \ H_0^{eff}=-\sum_{a=1}^n \frac{1}{4E_c}\sum_i\left( \frac{\partial}{\partial\theta_i^a}\right)^2
\end{eqnarray}
where $a$ is the index of replica and $i$ is the site label.

The only difference from Ref\cite{dp2} with respect to the cumulant expansion
is the dissipation term.  However, inspection of the free energy indicates
that this term can be combined with the second-to-last term by defining a new Q matrix,
\begin{eqnarray}\label{Qt}
          \sqrt{2 J(k) J(k')}Q_{\mu \nu}^{a b} (k,k',\tau,\tau')-\delta_{ab}\delta_{\mu\nu}\delta_{k+k',0}
         \frac{\eta}{\pi(\tau-\tau')^2}\rightarrow \sqrt{2 J(k) J(k')}\tilde{Q}_{\mu \nu}^{a b} (k,k',\tau,\tau')
\end{eqnarray}
The new free energy,
\beq
\left\langle \hat{T} \exp\left(2\int^{\beta}_{0}d\tau \sum_{a,k}\sqrt{J_0(k)}\Psi_{\mu}^a (k,\tau)S_{\mu}^a (-k,\tau) \right.\right.\nonumber\\
\left. \left. +\sum_{a,b,k,k'}\int^{\beta}_{0} d\tau \int^{\beta}_{0} d\tau'\left[ \sqrt{2 J(k) J(k')}\tilde{Q}_{\mu \nu}^{a b} (k,k',\tau,\tau')\right] S_{\mu}^a (k,\tau) S_{\nu}^b (k',\tau') \right) \right\rangle_0, 
\eeq
is identical with $Q$ replaced with $\tilde{Q}$.  Hence, the solution to $\tilde{Q}$ is the same as that for $Q$.  Once the transformation, Eq. (\ref{Qt}) is undone, we recover that the Ohmic dissipation term simply renormalizes the total coefficient of $|\omega|$, leaving the glass and metal intact, contrary to the claims of Ikeda\cite{ikeda}.

\acknowledgements We thank Nadya Mason for providing the experimental data shown in Fig. (\ref{mkdata}) and the NSF, Grant No. DMR-0305864 for providing partial support for this work.


\begin{thebibliography}{99}
\bibitem{ikeda}R. Ikeda, cond-mat/0510380 (see version 3  of this paper).  In this version and the previous version v2, the author uses the same equation to
``easily find the only physically meanful solution", one is $D\propto |\omega|$ , the other is $D\propto |\omega|^{1/2}$. Appendix B shows that only the $|\omega|$ solution is correct.
\end{thebibliography}
\end{document}